\documentclass[fleqn,usenatbib]{mnras}

\usepackage{newtxtext,newtxmath}

\usepackage[T1]{fontenc}

\DeclareRobustCommand{\VAN}[3]{#2}
\let\VANthebibliography\thebibliography
\def\thebibliography{\DeclareRobustCommand{\VAN}[3]{##3}\VANthebibliography}

\usepackage{graphicx}	
\usepackage{amsmath}	

\title[Multiwavelength observations of PSR~J1544$-$2555]{Multiwavelength observations of a new black-widow millisecond pulsar PSR~J1544$-$2555}

\author[S. Belmonte Díaz et al.]{S. Belmonte Díaz$^{1}$\thanks{E-mail: sergio.belmontediaz@manchester.ac.uk (SBD); tinn.thongmeearkom@postgrad.manchester.ac.uk (TT); adipol.phosrisom@manchester.ac.uk (AP)}\thanks{These authors contributed equally to this work.}, T. Thongmeearkom$^{1,2}$\footnotemark[1]\footnotemark[2], A. Phosrisom$^{1,2}$\footnotemark[1]\footnotemark[2], R. P. Breton$^{1}$, M. Burgay$^{3}$, C. J. Clark$^{4,5}$,\newauthor L. Nieder$^{4,5}$, M. G. F. Mayer$^{6,7}$, W. Becker$^{6,8}$, E. D. Barr$^{8}$, S. Buchner$^{9}$, K. K. Das$^{10}$, V. S. Dhillon$^{11,12}$, \newauthor O. G. Dodge$^{4,5,1}$, E. C. Ferrara$^{13,14,15}$, J.-M. Griessmeier$^{16,17}$, R. Karuppusamy$^{8}$, M. R. Kennedy$^{18}$, M. Kramer$^{8,1}$, \newauthor P. V. Padmanabh$^{4,5}$, J. A. Paice$^{19,1}$, A. C. Rodríguez$^{10}$ and B. W. Stappers$^{1}$
\\
$^{1}$Jodrell Bank Centre for Astrophysics, Department of Physics and Astronomy, The University of Manchester, Manchester M13 9PL, UK\\
$^{2}$National Astronomical Research Institute of Thailand, Don Kaeo, Mae Rim, Chiang Mai 50180, Thailand\\
$^{3}$INAF – Osservatorio Astronomico di Cagliari, Via della Scienza 5, I-09047 Selargius (CA), Italy\\
$^{4}$Max Planck Institute for Gravitational Physics (Albert Einstein Institute), D-30167 Hannover, Germany\\
$^{5}$Leibniz Universität Hannover, D-30167 Hannover, Germany\\
$^{6}$Max-Planck Institut für extraterrestrische Physik, Giessenbachstrasse, 85741 Garching, Germany\\
$^{7}$Dr. Karl Remeis-Sternwarte and Erlangen Centre for Astroparticle Physics, Friedrich-Alexander Universität Erlangen-Nürnberg,
Sternwartstrasse 7, 96049\\ Bamberg, Germany\\
$^{8}$Max-Planck-Institut für Radioastronomie, Auf dem Hügel 69, D-53121 Bonn, Germany\\
$^{9}$South African Radio Astronomy Observatory, 2 Fir Street, Cape Town 7925, South Africa\\
$^{10}$Department of Astronomy, California Institute of Technology, 1200 East California Blvd, Pasadena, CA, 91125, USA\\
$^{11}$Astrophysics Research Cluster, School of Mathematical and Physical Sciences, University of Sheffield, Sheffield S3 7RH, UK\\
$^{12}$Instituto de Astrofísica de Canarias, E-38205 La Laguna, Tenerife, Spain\\
$^{13}$Department of Astronomy, University of Maryland, College Park, MD, 20742, USA\\
$^{14}$Center for Research and Exploration in Space Science \& Technology II (CRESST II), NASA/GSFC, Greenbelt, MD 20771, USA\\
$^{15}$NASA Goddard Space Flight Center, Greenbelt, MD 20771, USA\\
$^{16}$LPC2E - Université d’Orléans / CNRS, F-45071 Orléans cedex 2, France\\
$^{17}$Observatoire Radioastronomique de Nançay (ORN), Observatoire de Paris, Université PSL, Univ Orléans, CNRS, F-18330 Nançay, France\\
$^{18}$School of Physics, Kane Building, University College Cork, Cork T12 K8AF, Ireland\\
$^{19}$Centre for Extragalactic Astronomy, Department of Physics, Durham University, South Road, Durham DH1 3LE, UK\\
}

\date{Accepted XXX. Received YYY; in original form ZZZ}

\pubyear{2025}

\begin{document}
\label{firstpage}
\pagerange{\pageref{firstpage}--\pageref{lastpage}}
\maketitle

\begin{abstract}
We report the discovery of a new black-widow millisecond pulsar, PSR~J1544$-$2555, associated with the \textit{Fermi}-LAT source 4FGL~J1544.2$-$2554. Optical, radio, and gamma-ray observations confirmed its nature as a compact spider binary system. Optical photometry from ULTRACAM revealed a $\sim$2.7-hour orbital period, guiding MeerKAT observations that detected $\sim$2.4-ms radio pulsations. Subsequent timing campaigns using the Murriyang Parkes Telescope, the Effelsberg 100-m Radio Telescope, and the Nançay Radio Telescope allowed us to obtain a preliminary timing solution, which enabled us to find gamma-ray pulsations. The final timing solution, spanning 16 years of \textit{Fermi}-LAT gamma-ray data, also displays orbital period variations typical of spider pulsars. X-ray observations from eROSITA indicate non-thermal emission, but the relatively low count rate prohibits the search for X-ray pulsations. Optical light curve modelling using \texttt{Icarus} suggests the asymmetry is best explained by a spot model, where uneven heating creates localised temperature variations on the companion. While the optical spectra we obtained are compatible with the physical properties we infer for the companion star, they were not of sufficient signal-to-noise to allow for radial velocity measurements, thus limiting constraints on the neutron star’s mass. The observed bluer colour near the light curve minimum suggests possible non-thermal emission from intra-binary shocks, supported by the presence of an X-ray source. This discovery exemplifies the proven capability of the \textit{Fermi}-LAT catalogue in identifying millisecond pulsar candidates and highlights the role of optical surveys in detecting variable sources suitable for radio follow-up.
\end{abstract}




\section{Introduction}

Millisecond pulsars (MSPs) differ from the regular pulsar population due to their fast spin periods and low spin-down rates \citep{Backer82,Alpar1982,Backer83}. MSPs are thought to be formed in binary systems through the so-called recycling scenario, where accretion from the companion star spins up the pulsar through mass transfer \citep{BHATTACHARYA1991}. At this stage, the system would be observed as a low-mass X-ray binary (LMXB). The system would return to emit radio pulsations once mass accretion ceases due to the disappearance of the accretion disc (e.g. after the companion no longer overfills its Roche lobe). Therefore, the recycling scenario suggests that there is a link between LMXBs and millisecond radio pulsars, which is evidenced by the discovery of accreting millisecond X-ray pulsars \citep{AccretingmilisecondXraypulsar1998}, followed by the detection of transitional MSPs \citep[tMSPSs;][]{Archibald2009,Papitto2013,Stappers2014}. tMSPs are binary systems in which a pulsar has been spun up enough to start emitting radio pulses. However, these systems can transition back to an accretion state, during which material from the companion star obscures or quenches the radio emission.

Spider pulsars are an emerging class of MSP binaries. They contain an MSP and a tidally-locked low-mass stellar companion in a compact orbit \citep[$P_{\rm orb} \lesssim$ 24 hours;][]{Roberts2011}.The companion mass divides the population into two sub-categories: black widows contain a lower mass companion ($\ll 0.1 {\rm M}_\odot$) in comparison to redbacks \citep[$ \gtrsim 0.1 {\rm M}_\odot$;][]{Roberts_2013}. These energetic systems can emit radiation across the entire electromagnetic spectrum. The MSP dominates the gamma-ray and radio emission. Optical emission is dominated by the companion star, which is tidally distorted due to the close orbit and strongly irradiated by the pulsar wind \citep{Djorgovski1988}. Variability can be observed at a combination of half the orbital period, due to ellipsoidal modulation from tidal distortion, and at the orbital period, due to changes in the visibility of the irradiated companion star when the temperature on the day side is substantially above that of the night side   . An intrabinary shock (IBS) is created when the relativistic pulsar wind collides with the companion stellar wind, which dominates the X-ray emission \citep{Arons1993}.

The multiwavelength nature of the emission produced by spiders can be used to facilitate their detection. While the presence of an MSP can be confirmed by finding radio or gamma-ray pulsations, discovering new binaries through radio observations is challenging. The tight orbit of these binary systems causes the pulsar to have a relatively large acceleration, hindering the search \citep{Johnston1991}. Moreover, the material ablated from the companion’s surface by the pulsar wind has the potential to block the pulsar's radio emissions, leading to eclipses lasting a substantial portion of the orbit \citep[e.g.][]{Fruchter1988,Lyne1990,Polzin2020}. An alternative approach is to search for steep-spectrum radio continuum sources. Previous work \citep{Frail2016} has shown that gamma-ray MSPs often exhibit unusually steep spectral indices ($\alpha \leq -2.5$), which may make them less affected by tight orbital motions or propagation effects, potentially offering a more reliable way to identify new MSP candidates. \citet{Frail2018} further demonstrated that identifying steep-spectrum sources can help guide deeper pulsation searches for MSPs.

So far, the most efficient way to find spiders has been through their gamma-ray emission. Since the \textit{Fermi Gamma-ray Space Telescope} started to operate in 2008, the Large Area Telescope \citep[LAT;][]{LAT2009} has detected gamma-ray pulsations of more than 300 pulsars \citep[see, e.g.][]{Smith2023,Thongmeearkom2024+RBs}. Roughly half are MSPs and half are young pulsars. Unlike young pulsars, MSPs tend to reside at higher Galactic latitudes due to their advanced age, which provides them with more time to migrate away from the Galactic plane. Traditional radio pulsar surveys have primarily focused on the Galactic plane \citep[e.g.][]{Manchester2001+ParkesSurvey,Keith2010+HTRUPS}, leaving high-latitude regions less explored. The wide field of view of Fermi and its ability to scan the entire sky every three hours make it particularly effective at identifying gamma-ray emission from pulsars over the whole sky. The spectral characteristics of gamma-ray emission from unidentified sources detected by Fermi provide a powerful means of identifying pulsar-like candidates, which can then guide targeted follow-up searches for pulsations with deep radio observations \citep[e.g.][]{Ray2012+PSC,Clark2023}. In addition, optical and X-ray observations provide further means of uncovering potential counterparts and characterising these systems \citep[e.g.][]{Lu2024+opticalsurvey,Mayer2024}. The fourth release of the \textit{Fermi}-LAT source catalogue \citep[4FGL-DR4;][]{Abdo2020, 4FGLDR4} contains more than 7,000 sources, of which around 2,600 remain unidentified\footnote{\url{https://fermi.gsfc.nasa.gov/ssc/data/access/lat/14yr_catalog/}}. While many of these are likely Active Galactic Nuclei, they still offer a promising pool for potential pulsar discoveries.

One important property of spider pulsars is their tendency to host heavier pulsars. Based on a sample of 15 objects, \citet{Strader2019} reported a median mass of $1.78 \pm 0.09\,{\rm M}_\odot$ in redback systems, which is significantly larger than the canonical neutron star mass of $1.4\,{\rm M}_\odot$. High pulsar masses help constrain the equation of state of multiple theoretical models, providing useful insights into the properties of cold, ultradense matter \citep[see][]{Ozel2016}. Some systems have been found to possibly host a neutron star of mass $\gtrsim 2.1\,{\rm M}_\odot$, such as the black widows PSR~J0952$-$0607 \citep[$2.35 \pm 0.17\,{\rm M}_\odot$;][]{Romani2022}, PSR~J1653$-$0158 \citep[$2.17 \pm 0.2\,{\rm M}_\odot$;][]{Nieder2020}, and PSR~J1810+1744 \citep[$2.13 \pm 0.04\,{\rm M}_\odot$;][]{Romani2021}, and the extremely irradiated redback PSR~J2215+5135 \citep[$2.27 \pm 0.16\,{\rm M}_\odot$;][]{Linares2018}. The current limitation on these mass measurements is that they rely on the light curve and spectral modelling to infer the orbital inclination and companion's semi-major axis, where systematic uncertainties are still difficult to ascertain especially when light curves present asymmetries. \citet{Clark2023b} recently used gamma-ray eclipses to independently constrain the inclination of spider binaries, which resulted in a reduction in mass for the pulsar PSR~B1957+20 from $2.40 \pm 0.12\,{\rm M}_\odot$ \citep{vanKerkwijk2011} to $1.81 \pm 0.07\,{\rm M}_\odot$ \citep[see also the case for PSR~J1311$-$3430 in][]{Romani2015}.

In this paper, we present the discovery of a new black-widow MSP associated with the \textit{Fermi}-LAT source 4FGL~J1544.2$-$2554. A co-located source showing optical variability of $\sim$ 2.7 hours was found from multi-band light curves obtained from ULTRACAM observations. The light curve shows properties that are in agreement with black-widow systems. The nature of the source was confirmed with the discovery of $\sim$ 2.4 ms radio pulsations from MeerKAT observations. Independently from this work, \citet{Karpova2024+1544} proposed that 4FGL~J1544.2$-$2554 is a spider candidate with period 2.7 hours based on a single-peak optical light curve from the 2.1-m OAN-SPM telescope. In addition, this source was considered a pulsar-like candidate from X-ray emission detected with the eROSITA telescope \citep{Mayer2024}. Some previously identified black-widow systems were first discovered through optical observations and subsequently confirmed via gamma-ray pulsation searches \citep[see][]{Pletsch2012, Nieder2020}. However, to our knowledge, this is the first black-widow system to be confirmed through radio follow-up observations after being initially identified as an optical candidate. Notably, redback systems PSR~J2339$-$0533 \citep{Romani2011,Kong2012,Ray2020}, PSR~J0212+5320 \citep{Li2016,Linares2017,Perez2023}, and the three pulsars discovered in \citet{Thongmeearkom2024+RBs} followed a similar discovery path. Searching for black widows through optical observations is challenging because they are typically faint sources, requiring the use of highly sensitive telescopes.

\section{Optical discovery}
\label{S:Optical_discovery}
The \textit{Fermi}-LAT unidentified source 4FGL~J1544.2$-$2554 was deemed a pulsar candidate based on its gamma-ray spectral properties. Pulsars tend to show low long-term gamma-ray variability and a curved spectrum that deviates from a single power-law at high photon energies \citep{Abdo_2010}. These features distinguish them from the other main class of gamma-ray sources: Active Galactic Nuclei, which show variability over time and have a flatter spectrum \citep{SazParkinson_2016}.

The field was observed as part of a survey we conducted to search for periodic optical sources in 35 \textit{Fermi}-LAT unidentified sources (Phosrisom et al. in preparation) using ULTRACAM \citep{ULTRACAM}, a high-speed imaging photometer mounted at the 3.5-m New Technology Telescope at the European Southern Observatory in Chile \citep{NTT}.
The observing area was centred at the middle of the \textit{Fermi}-LAT localisation ellipse containing the 95$\%$ confidence level of the source's location. The field was observed at Super Sloan Digital Sky Survey bands \textit{u$_{s}$,g$_{s}$,r$_{s}$} simultaneously, with exposure times of approximately 20 seconds, and a dead time of only 24 milliseconds between each exposure. Observations were conducted for a total of 9 hours between April and June 2022, with an additional two hours and 30 minutes of observation in February 2024, as shown in Table \ref{tab:optical_observing_hours}.  Data were recorded under photometric conditions, with atmospheric seeing fluctuating from 1 to 1.5 arcseconds. Observations were taken at different sidereal times and spread over several weeks to reduce potential aliasing effects when searching for periodic signals. The final set of observations, conducted long after the initial epochs, was used to constrain the orbital period ($P_{\rm orb}$) of potential candidates.

\begin{table}
	\centering
	\caption{Time and orbital phase coverage of the ULTRACAM observations of 4FGL~J1544.2$-$2554. The orbital phase coverage was obtained using the timing solution reported in Table \ref{tab:timing_1544}.}
	\label{tab:optical_observing_hours}
	\begin{tabular}{ccc} 
		\hline
		Start time & Observation length & Orbital phase \\
            (UTC)      & (min)              & coverage       \\
		\hline
        \hline
		03/04/2022 06:00:00 & 60.0 & 0.38 $-$ 0.74\\
		04/04/2022 05:04:00 & 56.0 & 0.84 $-$ 1.19\\
        05/04/2022 05:02:00 & 58.0 & 0.64 $-$ 0.99\\
		06/04/2022 05:09:49 & 50.5 & 0.50 $-$ 0.81\\
        07/04/2022 05:02:42 & 58.5 & 0.27 $-$ 0.62\\
        29/04/2022 03:59:54 & 58.3 & 0.73 $-$ 1.08\\
        30/06/2022 03:02:28 & 179.3 & 0.65 $-$ 1.75\\
        11/02/2024 06:17:15 & 149.2 & 0.12 $-$ 1.03\\
		\hline
	\end{tabular}
\end{table}

The HiPERCAM \citep{HiPERCAM} pipeline\footnote{\url{https://cygnus.astro.warwick.ac.uk/phsaap/hipercam/docs/html/}} was used to reduce the data. A Lomb-Scargle periodogram \citep{Lomb, Scargle} was used to search for periodic variability ranging from one to 48 hours. This is the period range at which spiders are expected to be detected in optical frequencies. In addition to analysing the best two periods of the search, a fold was done at double and half their values to ensure the fundamental period was found. This is motivated by spider pulsar systems showing both single and double-peak light curves. Each source was individually inspected to assess its nature.

\begin{figure*}
    \centering
	\includegraphics[width= \textwidth]{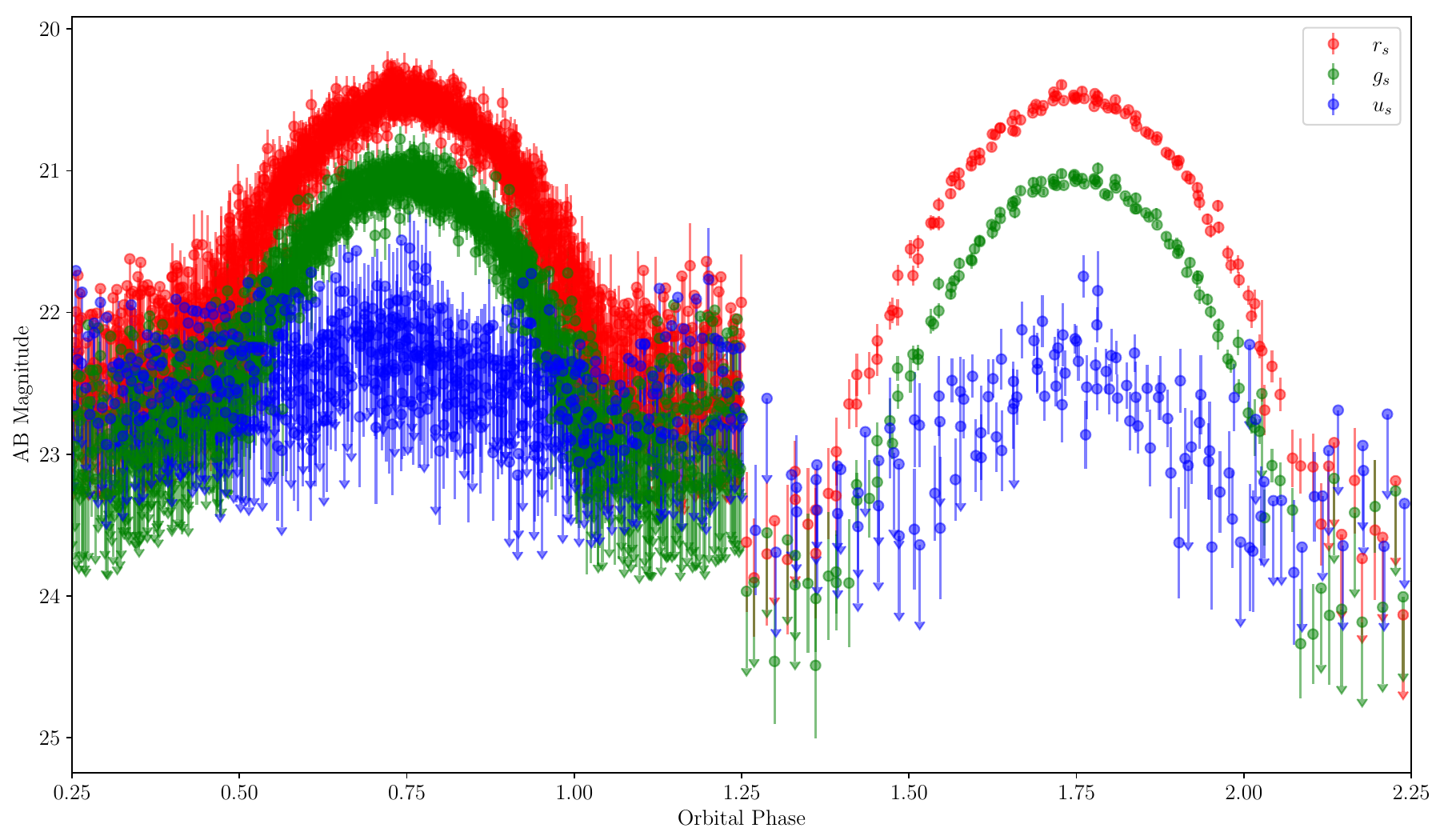}
    \caption{ULTRACAM optical light curve of the periodic source in the $r_{s}$, $g_{s}$ and $u_{s}$ bands folded at the best orbital period found from the Lomb-Scargle periodogram (0.113495 days). In the first cycle (left), all available data points are plotted and in the second cycle (right) data points are averaged to ease the visualisation of the lightcurve's trend. Orbital phases are defined such as 0.25 corresponds to the pulsar's superior conjunction.}
    \label{fig:lightcurve_mag}
\end{figure*}

\begin{figure}
    \centering
	\includegraphics[width= \columnwidth]{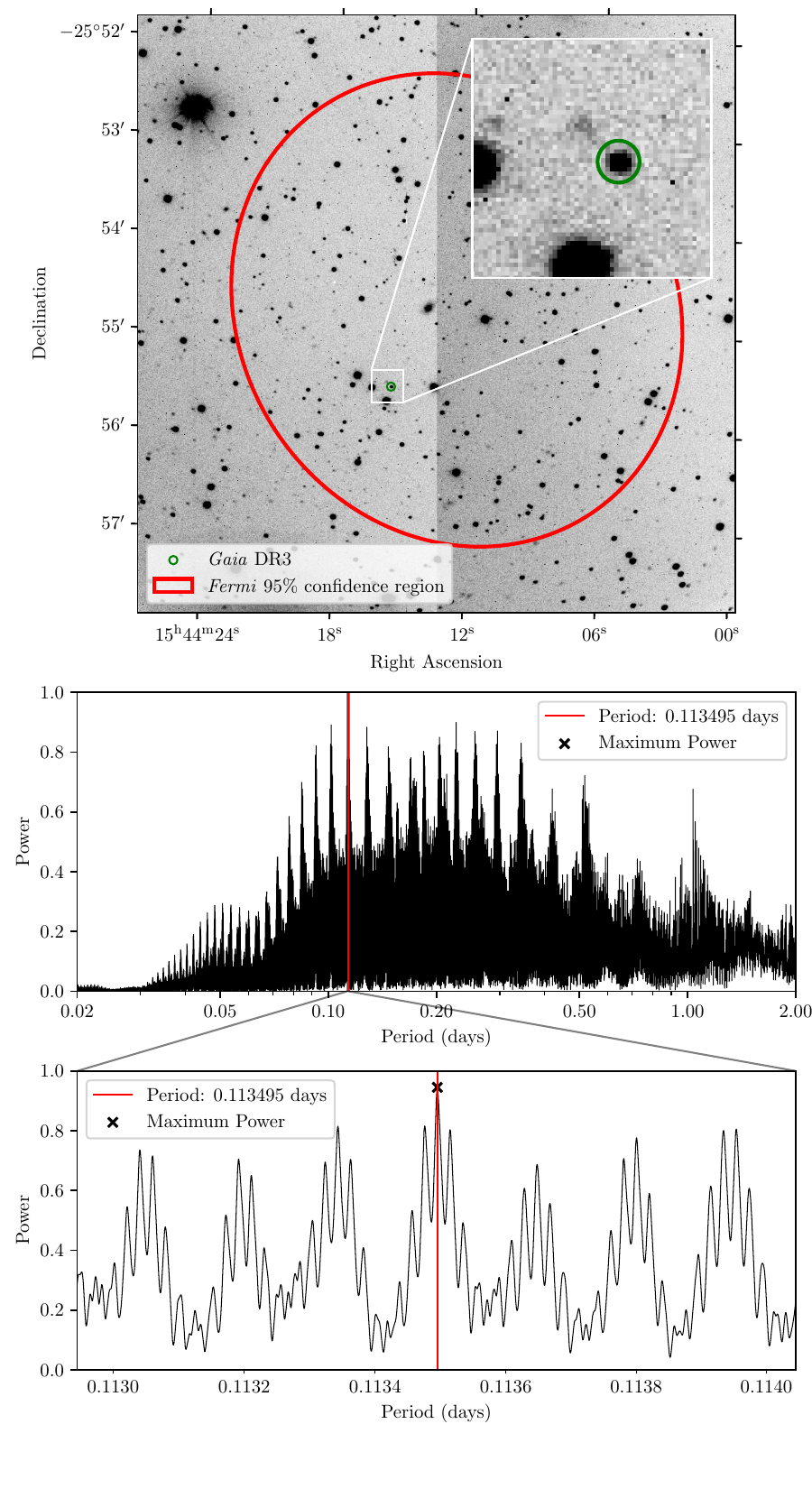}
    \caption{Top: finding chart showing the location of the periodic variable source in the ULTRACAM observing field. The \textit{Fermi}-LAT localisation ellipse containing the 95$\%$ confidence level of the location of the source is shown in red. The location of the Gaia counterpart is encircled in green. Middle: Full Lomb-Scargle periodogram computed using the data recorded from the optical observations, shown over the period range 0.02 - 2 days. The red vertical line shows the best recovered period. Bottom: Zoomed-in view of the Lomb-Scargle periodogram around the best-recovered period. The red vertical line shows the best recovered period, and the black cross shows the maximum power at the best orbital period.}
    \label{fig:loc_and_ls}
\end{figure}

A source was found to show a single-peak light curve with a period of $\sim$ 2.7 hours (Figure~\ref{fig:lightcurve_mag}). The values of the orbital period and large optical variability ($\gtrsim 3$\,mag) presenting a single peak per cycle are in agreement with the candidate being a black widow pulsar binary \citep[e.g.][]{Mata-Sanchez2023}. A \textit{Gaia} counterpart was also found and provides an accurate position, beneficial for eventual radio timing. A finder chart with the localisation of the source and a Lomb-Scargle periodogram are shown in Figure \ref{fig:loc_and_ls}. The long baseline of 1.85 years between observations helped reduce aliasing in the periodogram and provided an accurate orbital period $P_{\rm orb} = 0.11349520(4)$ days (3\,ms uncertainty). Modelling the light curve with a simple two-term sinusoid, $A_{0}+A_{1}\sin{(2\pi(t-T_{\rm asc})/P_{\rm orb}+\pi)}+A_{2}\cos{(4\pi(t-T_{\rm asc})/P_{\rm orb})}$, representing the mean flux, the irradiation and the ellipsoidal variability enables ones to determine the epoch of ascending node of the unseen compact object\footnote{This is to follow orbital phase convention used for pulsar timing.} $T_{\rm asc} = {\rm MJD}~59672.32042(7)$ (6\,s uncertainty). This statistical uncertainty reflects the fit of a simple symmetric heating model to the light curve. However, deviations from a perfectly symmetric heating distribution could introduce bias in $T_{\rm asc}$ (see \S\ref{S:Optical_modelling} for evidence of a slight asymmetry in the light curve).

4FGL J1544.2-2554 was also observed with the Keck I telescope using the Low-Resolution Imaging Spectrometer \citep[LRIS;][]{1995lris} on 08 February 2024 and 15 February 2024 (UT). On the first night, we used the 600/4000 grism on the blue side with 2x2 binning (spatial, spectral), and the 400/8500 grating on the red side with 2x1 binning. On the second night, we used the 400/3400 grism on the blue side with 2x2 binning (spatial, spectral), and the 400/8500 grating on the red side with 2x1 binning. On both occasions, we used a 1.0$\arcsec$ slit, and the seeing each night was approximately 0.7--1$\arcsec$, leading to minimal slit losses. All Keck I/LRIS data were reduced with \texttt{lpipe}, an IDL-based pipeline optimized for LRIS long slit spectroscopy and imaging \citep{2019perley_lpipe}. Data were flat fielded sky-subtracted, and heliocentric corrected using standard techniques. Internal arc lamps were used for the wavelength calibration and a standard star for overall flux calibration.

\section{Discovery of radio pulsations}

Given the evidence provided by the optical light curve from the ULTRACAM observations and by the gamma-ray spectrum from \textit{Fermi}-LAT, there is a high likelihood that 4FGL~J1544.2$-$2554 contains a spider pulsar binary. We therefore decided to observe this field with the MeerKAT telescope and search for radio pulsations, as a detection would confirm the presence of a MSP and elucidate the true nature of the gamma-ray point source.

MeerKAT is an interferometer array telescope located in the Karoo, in the Northern Cape of South Africa. Locating the telescope in a remote area reduces radio frequency interference (RFI) to low levels, which makes the MeerKAT telescope a great tool for radio astronomy. It contains 64 steerable radio antennas with a diameter of 13.5m. All antennas are equipped with S-band (1750-3500 MHz), L-band (856-1712 MHz), and Ultra High Frequency (UHF) (544-1088 MHz) receivers, offering a wide range of frequency coverage \citep{Jonas_2016}. MeerKAT provides an increase in sensitivity concerning previous radio telescopes in the southern hemisphere, allowing us to find fainter pulsars. 

\subsection{Observation configuration}
We observed the candidate at UHF since pulsars often show steep radio spectra, increasing their detectability at low-frequencies \citep{Jankowski2017}. The optical ephemeris provided an approximate orbital period and reference orbital phase, which were used to schedule the observation around the inferior conjunction of the pulsar, where radio eclipses are less likely to occur as the companion is located behind. Following the survey configuration in \citet{Thongmeearkom2024+RBs}, we scheduled a one-hour radio observation that covers $\sim 35 \%$ of the orbit. The extent of orbital coverage enables the system to be observed predominantly during non-eclipsing orbital phases, even if the orbital ephemeris deviates slightly from the actual parameters. At the time of the observation, 60 MeerKAT dishes were available. Since the optical counterpart offers a highly accurate localisation, a single tied-array beam was used. The data were recorded at a sampling time of 60\,$\mu$s and 4096 frequency channels on the 7th of June of 2023 (MJD~60,102).

\subsection{Pulsation searches}
We performed the search using \texttt{peasoup}\footnote{\url{https://github.com/ewanbarr/peasoup}} \citep{Barr2020+peasoup}, a GPU-accelerated code that performs a time-domain resampling acceleration search. The software implements a dedispersion step, dereddening, and an FFT-based acceleration search through time-domain resampling with incoherent harmonic summing \citep[described in detail in][]{Morellopeasoup}.

We search for radio pulsations over a dispersion measure (DM) range of $0-180$\,pc\,cm$^{-3}$, with a DM step of 0.1\,pc\,cm$^{-3}$; and over accelerations up to $50$\,m\,s$^{-2}$. The DM range was chosen by considering twice the maximum amount of dispersion predicted from the electron density models YMW16 \citep{Yao_2017} and NE2001 \citep{Cordes2001} to ensure the real DM is not missed. The acceleration limit was chosen considering the maximum acceleration a spider pulsar can experience given an approximate orbital period, pulsar mass, and companion mass typical of these systems. We coherently searched in the full-hour observation and in 30-minute and 15-minute segments. The search in the shorter segments was motivated by the fact that shorter observing lengths enhance the detection of pulsars in tight orbits, where optimal observation times for MSPs with orbits of less than a day range from 5$\%$ to 20$\%$ of the orbital period \citep{Johnston1991}.

We initially found a candidate with a period of $\sim$ 2.4 ms in one of the 15-minute segments. Further analysis of the full hour revealed it in other segments. The short period and the large acceleration ($\sim$ 15.8 m s$^{-2}$) confirms the binary nature of this new MSP.

\begin{table}
  \centering
  \caption{Timing observation table containing the number of observations (N. obs), the total duration of the observations (tobs), and the number of TOAs (N. TOAs) used to obtain the radio timing solution. The information is provided for the different instruments used to observe the source: MeerKAT (MKT), Parkes (PKS), Nan\c{c}ay (NRT), and Effelsberg (EFF). Observations with weak detections, non-detections, and the last Parkes detection (see Figure \ref{fig:phase_cover}) are not included in the total counts since they were not used to obtain the timing solution.}
  \label{tab:timing_obs}
  \begin{tabular}{cccc}
    \hline
    Telescope & N. obs & tobs (h) & N. TOAs \\
    \hline
    MKT & 1 & 1 & 60 \\
    PKS & 3 & 3 & 26 \\
    NRT & 2 & 2 & 14 \\
    EFF & 1 & 1 & 2\\
    \hline
  \end{tabular}
\end{table}

\begin{figure}
    \centering
	\includegraphics[width=\columnwidth]{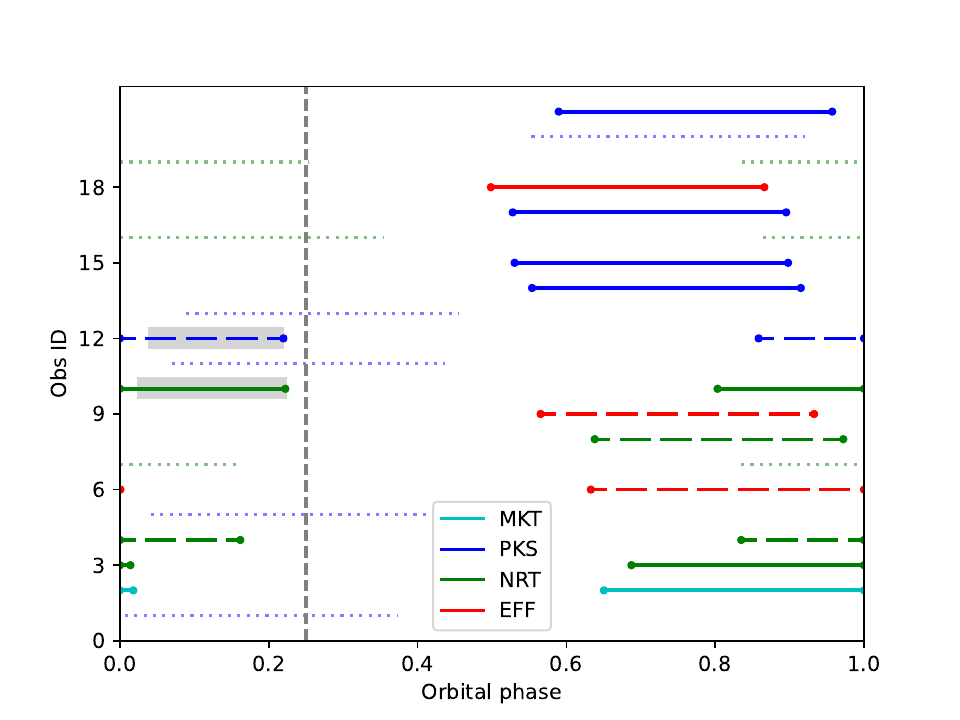}
    \caption{Orbital phase coverage of radio observations obtained with different telescopes: MeerKAT (MKT) in cyan, Parkes (PKS) in blue, Nan\c{c}ay (NRT) in green, and Effelsberg (EFF) in red. Phase 0 corresponds to the pulsar at the ascending node, and phase 0.25 corresponds to the pulsar's superior conjunction. Strong detections are displayed by brighter solid lines, while weak detections and non-detections are displayed with a dimmer dashed and dotted line respectively. Only strong detections were used for pulsar timing (Table \ref{tab:timing_obs}). The grey-shaded region indicates when a clear eclipse was detected in the observation.}
    \label{fig:phase_cover}
\end{figure}

\begin{figure}
    \centering
	\includegraphics[width=\columnwidth]{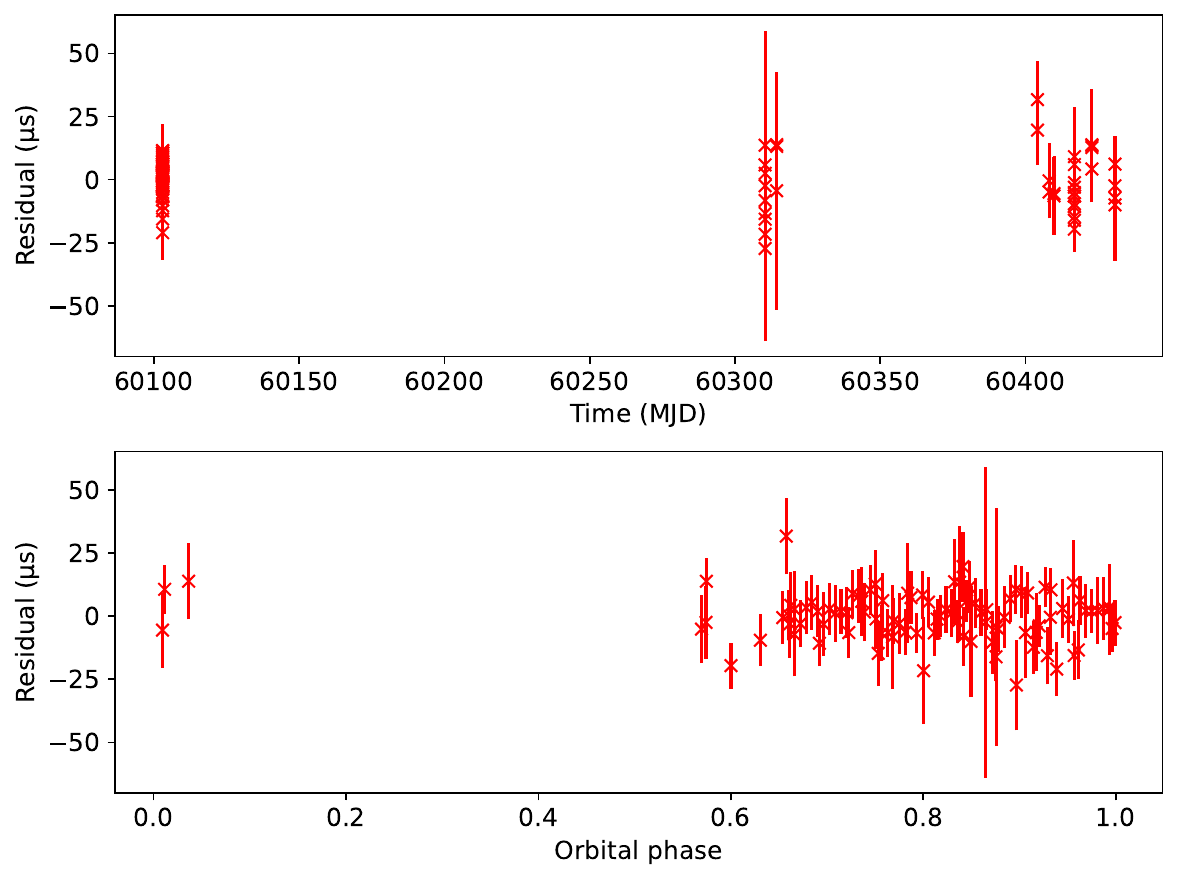}
    \caption{Radio residuals for PSR~J1544$-$2555 using the best ephemeris obtained in the joint radio and gamma-ray timing analysis (Table~\ref{tab:timing_1544}), shown versus time (top panel) and orbital phase (bottom panel).}
    \label{fig:residuals}
\end{figure}

\subsection{Radio timing}
\label{sec:radio_timing}
We initiated a timing campaign for PSR~J1544$-$2555 using the Murriyang Parkes telescope\footnote{\url{https://www.csiro.au/en/about/facilities-collections/ATNF/Parkes-radio-telescope-Murriyang}}, the Effelsberg 100-m radio telescope\footnote{\url{https://eff100mwiki.mpifr-bonn.mpg.de/doku.php}}, and the Nançay Radio Telescope \citep[NRT;][]{NRT}. Timing of a pulsar binary system involves fitting the pulse times of arrival (TOAs) with a model parametrised by the spin period ($P$), dispersion measure (DM), orbital period ($P_{\rm orb}$), epoch of ascending node ($T_{\rm asc}$), projected semi-major axis ($x$) and the source's sky position.

The discovery time-vs-spin-phase plot showed a curved pulsar signal, indicating that the folding parameters found by the acceleration search software were not optimal. To refine these parameters, we measured the pulsar's spin period across the one-hour MeerKAT observation using \texttt{PREPFOLD} from \texttt{PRESTO}\footnote{\url{https://github.com/scottransom/presto}} \citep{Ransom2011+PRESTO} on five-minute segments. The short segment duration allowed accurate determination of the spin period at different orbital phases. The barycentric spin periods obtained were fitted for a circular orbit using  \texttt{fit\_circular\_orbit.py} from \texttt{PRESTO}, yielding improved values for $x$, $P_{\rm orb}$, and $T_{\rm asc}$. However, we found that the estimate of $P_{\rm orb}$ derived from the optical periodogram (see \S\ref{S:Optical_discovery}) was more accurate than the estimate from \texttt{fit\_circular\_orbit.py}, due to the limited initial radio detections and the longer baseline of optical observations. Therefore, the $P_{\rm orb}$ from the optical periodogram and the $x$ value from \texttt{fit\_circular\_orbit.py} were used as starting parameters for timing follow-up observations with \texttt{tempo2} \citep{tempo2}. The sky position was fixed to the value from \textit{Gaia}, the DM to the value maximising the S/N in the full MeerKAT observation, and $P_{\rm orb}$ to the optical periodogram value. To account for time delays between observations with different telescopes, a "JUMP" parameter was included in the timing model, allowing for spin-phase adjustments between TOAs from different instruments. Table \ref{tab:timing_obs} shows the number of observations, observing time, and number of TOAs for each telescope. The number of TOAs depended on the brightness of the observation. Note that we did not include TOAs from observations in which the pulsar was very weak (with a \texttt{PRESTO} S/N below 6) as they tend to produce poor results. Figure \ref{fig:phase_cover} illustrates the orbital phase coverage of the observations, with non-detections and weak detections highlighted by different colour markers and line styles. We managed to connect the MeerKAT discovery observation from June 2023 to the latest Effelsberg observation in May 2024 (see the radio residuals in Figure~\ref{fig:residuals}). The timing solution for PSR~J1544$-$2555 is shown in Table \ref{tab:timing_1544}.

Spider systems are known to exhibit long-term orbital period variations \citep[e.g.][]{Arzoumanian1994+orbitalmodulation,Voisin2020+orbitalmodulation,Thongmeearkom2024+RBs} (see also \S\ref{sec:gamma-ray_pulsation}). High orbital frequency derivatives can reduce the S/N of the folded pulse due to small changes in the orbital period. This effect can be mitigated by shifting the reference spin phase iteratively until the signal is recovered. To investigate whether the non-detections were caused by significant orbital frequency variations, we used \texttt{SPIDER$\_$TWISTER}\footnote{\url{https://github.com/alex88ridolfi/SPIDER_TWISTER}}, a Python-based tool designed to fold pulsar data across a range of $T_{\rm asc}$ values, allowing for searches across multiple orbital phases. Since no significant pulsations were detected, this suggests that the non-detections are unlikely to be caused by high orbital frequency derivatives. Instead, they are more consistent with eclipses. Further supporting this interpretation is the fact that our radio timing solution is sufficiently precise for gamma-ray timing, as evidenced by the detection of gamma-ray pulsations spanning 16 years of \textit{Fermi} data (see \S\ref{sec:gamma-ray_pulsation}).

We found a 2016 archival radio observation of this Fermi field with the Parkes telescope \citep[project P814][]{Camilo2017+arxobs}. The sky position of this observation is about 1 arcminute away from the position of the pulsar and there comfortably falls within the primary beam of the telescope. We folded these data using the final gamma-ray timing solution (see \ref{sec:gamma-ray_pulsation}) to search for radio pulsations, but none were detected. The orbital phase coverage of the observation spans from phase 0 to 0.4, corresponding to the eclipsing region based on other observations (Figure \ref{fig:phase_cover}). A recent study has analysed this observation in detail \citep{Kerr2025}, but no pulsations were reported there either.

\section{Gamma-ray pulsations}
\label{sec:gamma-ray_pulsation}
To search for gamma-ray pulsations from this pulsar, we used data recorded by the \textit{Fermi} Large Area Telescope (LAT). We used \texttt{SOURCE}-class photons according to the `Pass 8' \texttt{P8R3\_SOURCE\_V3} instrument response functions \citep{Atwood2013+Pass8,Bruel2018+Pass8} and selected photon energies similar to the 4FGL-DR4 source catalog \citep{4FGLDR4}. Each gamma-ray photon is assigned a photon weight, representing the probability the photon is from the foreground compared to the background \citep{Bickel2008+weights,Kerr2011+weights,Bruel2019+weights}. This is computed using spectral and spatial models of the gamma-ray sky using \texttt{gtsrcprob}. The sources’ positions and spectra are taken from 4FGL-DR4 \citep{4FGLDR4}, with the Galactic and diffuse emission models, \texttt{gll\_iem\_uw1216\_v13.fits} and \texttt{iso\_P8R3\_SOURCE\_V3\_v1.txt}.

To improve computing efficiency in pulsation search and timing analysis, we performed a cut on the photon weights. The pulsation significance depends directly on these weights $w_j$ and scales linearly with $W^2 = \sum_j w_j^2$. The expected value for the widely used weighted $H$-test statistic \citep{Kerr2011+weights} is proportional to $W^2$ \citep{Nieder2020+metric}, with a source-dependent prefactor describing pulse profile and background, which is typically of order unity. The scaling is used to remove the majority of low-weight photons by introducing a minimum weight $w_{\rm min}$ which still retains $99\%$ of the signal. For this source, we chose $w_{\rm min} = 0.01$ which reduced the number of photons by $90\%$ while preserving $W^2 = 380$.

The gamma-ray pulsation search was guided by the pulsar parameter values as measured in the radio timing analysis and for the likely optical counterpart. The astrometric parameters are provided to high enough precision in \textit{Gaia} DR3 \citep{Gaia+DR3} that a search over these was not required. Radio timing allowed us to fix two further parameters, the time of ascending node $T_{\rm asc}$ and projected semi-major axis $x$, and to search two other parameters within their uncertainty margins, spin frequency $f$ and orbital period $P_{\rm orb}$. As radio timing is usually more sensitive to eccentric orbits than gamma-ray searches and did not show signs of non-negligible eccentricity, we assumed a circular orbit in the search. We searched the spin-frequency derivative $\dot{f}$ in the range $0$ to $-2 \times 10^{-14}$\,Hz/s, which is guided by the population of known MSPs.

The search was performed on a single compute node with a GPU, assuming a constant orbital period. The code searches over the $f$ range using fast Fourier transforms (FFTs), while looping over a lattice of $P_{\rm orb}$ and $\dot{f}$ trials. The lattice is built using a parameter space metric \citep{Nieder2020+metric}, allowing a maximum expected signal loss of $5\%$ in the third harmonic.

The search found a $H$ statistic maximum of $H=76.3$, which is statistically very significant, but much lower than expected for a pulsar with $W^2=380$. This could be explained by pulsar properties not accounted for by our spin-phase model. Over the course of the gamma-ray data span, pulsations in a time-vs-spin-phase plot seem to come into and out of focus several times which is typical for spider pulsars where variations of the orbital period have not been accounted for.

Quasi-random variability around a central orbital period is common among spider pulsars. These orbital period variations (OPVs) have been seen for nearly every redback pulsar \citep[e.g.][]{Clark2021+j2039,Thongmeearkom2024+RBs,Burgay2024+timing,Corcoran2024+GCredbacks}, and several black widow pulsars \citep[e.g.][]{Voisin2020+GQM}. This behaviour is usually thought to be a consequence of gravitational quadrupole moment variations of the pulsar's companion star \citep[e.g.][]{Applegate1994+B1957,Lazaridis2011+J2051}.

\begin{figure*}
    \centering
    \includegraphics[width=0.8\textwidth]{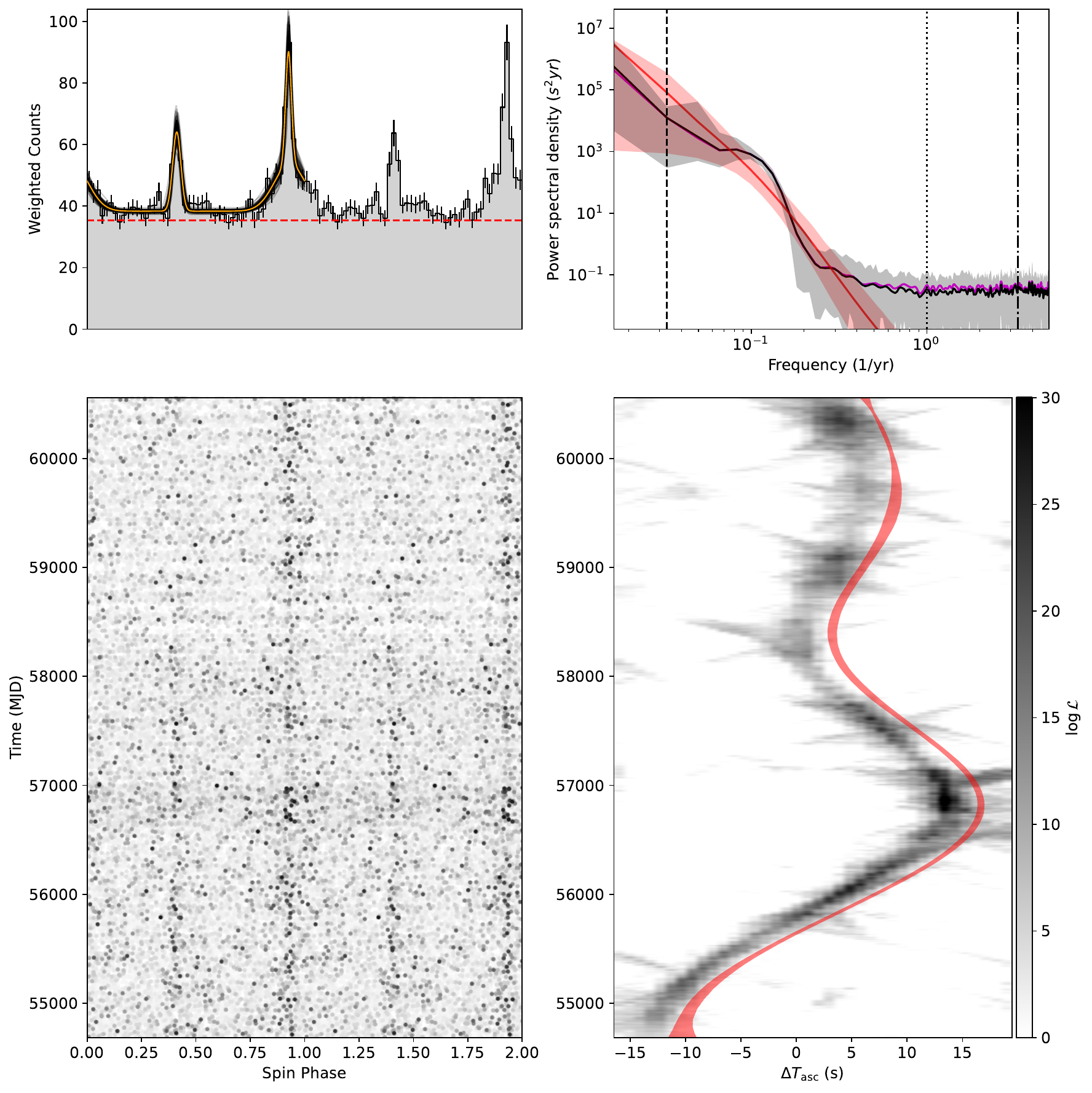}
    \caption{Gamma-ray pulsations and variations of the orbital phase over time for PSR~J1544$-$2555. The panels on the left show the weighted pulsar spin phases for each gamma-ray photon for the highest-likelihood timing solution (lower panel) and the integrated pulse profile (upper panel). The orange curve represents the associated highest-likelihood pulse-profile template and the 100 faint black curves are randomly drawn samples from the Monte-Carlo analysis to visualize the uncertainty on the pulse-profile template. The bottom-right panel shows the orbital phase variations as a function of time. The grey-scale image shows the log-likelihood for offsets from the pulsar's $T_{\rm asc}$ as measured in overlapping $400$-day windows. The red curves represent the $95\%$ confidence interval on those deviations, obtained from Monte-Carlo timing analysis, with a $3$\,s-offset for clarity. The top-right panel shows the power spectral density of the orbital phase variations. The red shaded region illustrates the 95\% confidence interval of the Mat\'{e}rn model. The black curve shows the estimated power spectrum from the joint radio and gamma-ray fitting and the grey shaded region its 95\% confidence interval. The purple curve, mostly hidden behind the black one, shows the power spectrum estimated from gamma-ray timing alone.}
    \label{f:J1544_fermi_timing}
\end{figure*}

The first stage of our timing code is using the sliding-window technique to provide a first model tracking the orbital period variations over time. This technique slides a window of a couple of hundred days length over the 16 years of data and searches within these time spans over time of ascending node and orbital period offsets from the nominal central value\footnote{OPVs on shorter time scales than the window span would necessitate to search orbital frequency derivatives as well}. In this case, a 400-day window has been sufficient, being short enough that the assumption of a constant orbital period is acceptable and being long enough to accumulate sufficient S/N. The result is shown in gray scale in Fig.~\ref{f:J1544_fermi_timing}, showing the cumulative long-term variation of the orbital period in the ascending node if compared to a constant orbital period model. This is used to fit an initial Gaussian process model.

The second stage uses the OPV model from the sliding-window method as a starting point, and then jointly refines it along with the other pulsar parameters and the gamma-ray pulse template, using both gamma-ray photons and radio TOAs. We also fit three OPV ``hyperparameters'' describing the covariance function of the orbital phase variations. For the latter, we apply the commonly-used Mat\'{e}rn covariance function \citep[e.g.][]{Rasmussen2006+GP}. It is described by an amplitude, $h$, length scale, $\ell$, and a smoothness parameter, $\nu$. This covariance function models the noise power spectral density by a smoothly broken power law, being flat below a corner frequency $f_{\rm c} = \sqrt{\nu}/\sqrt{2}\pi \ell$ and breaking to a power-law at higher frequencies with a spectral index of $\Gamma = -(2\nu + 1)$. We use a Gibbs sampling approach to deal with the multi-component likelihood function for each photon \citep{Thongmeearkom2024+RBs}. We will present more details on these methods and a release of the code in future works which are in preparation.

The timing analysis improved the $H$ statistic to $H=627.8$ with clear pulsations over the $16$ years of the \textit{Fermi} mission and a pulse profile with two narrow peaks, roughly half a rotation apart (see Fig.~\ref{f:J1544_fermi_timing}). The resulting pulsar parameters and estimated properties are listed in Table~\ref{tab:timing_1544}.

\begin{table}
  \centering
  \caption{Timing solution for the new black widow PSR~J1544$-$2555 using joint radio and gamma-ray timing with \textit{Gaia} positional parameters as priors. The derived parameters were obtained following the standard formulae \citep[e.g.][]{Smith2023}, assuming the YMW16 distance. The derived pulsar energetics are corrected for the Shklovskii effect \citep{Shklovskii1970+Pdot} and Galactic acceleration \citep[e.g.][]{Damour1991+Pdot}.}
  \label{tab:timing_1544}
  \begin{tabular}{lr}
    \hline
    Parameter & Value \\
    \hline
    \multicolumn{2}{c}{\textit{Gaia} DR3 astrometry}\\
    \hline
    R.A., $\alpha$ (J2000) & $15^h44^m15\fs4547(1)$\\
    Decl., $\delta$ (J2000) & $-25\degr55\arcmin32\farcs688(2)$\\
    Epoch of position measurement (MJD) & $57388.0$ \\
    \hline
    \multicolumn{2}{c}{Timing parameters}\\
    \hline
    Solar-system ephemeris & DE421\\
    Time scale & TDB \\
    Data span (MJD) & $54683 - 60560$\\
    Epoch of spin period measurement (MJD) & $60103.0$\\
    Proper motion in $\alpha$, $\mu_{\alpha} \cos \delta$ (mas yr$^{-1}$) & $-5.27\pm 0.84$\\
    Proper motion in $\delta$, $\mu_{\delta}$ (mas yr$^{-1}$) & $-18.1\pm 2.8$\\
    Spin frequency, $f$ (Hz) & $418.39828661948(4)$\\
    Spin-down rate, $\dot{f}$ (Hz s$^{-1}$) & $-2.0383(2)\times10^{-15}$\\
    Dispersion measure, DM (pc cm$^{-3}$) & $25.817 \pm 0.060$\\
    Orbital period, $P_{\rm orb}$ (d) & $0.113495141(9)$\\
    Projected semi-major axis, $x$ (lt-s) & $0.128832(5)$\\
    Epoch of ascending node, $T_{\rm asc}$ (MJD) & $56667.87739(9)$\\
    \hline
    \multicolumn{2}{c}{Hyperparameters of orbital-phase-covariance function}\\
    \hline
    Amplitude of OPV, $h$ (s) & $2^{+2}_{-1}$\\[0.2em]
    Length scale of OPV, $\ell$ (d) & $1920^{+4510}_{-670}$\\[0.2em]
    Smoothness parameter, $\nu$ & $>2.6$\\
    \hline
    \multicolumn{2}{c}{Derived parameters}\\
    \hline
    Distance, $d$ (kpc), NE2001 & $1.06$ \\
    Distance, $d$ (kpc), YMW16 & $1.02$ \\
    Spin period, $P$ (ms) & $2.3900671488879(3)$\\
    Spin period derivative, $\dot{P}$ & $1.1645(1)\times10^{-20}$\\
    Spin-down power, $\dot{E}$ (erg s$^{-1}$) & $2.8\times10^{34}$\\
    Surface magnetic field strength, $B_{\rm S}$ (G) & $1.5\times10^8$\\
    Light-cylinder magn. field strength, $B_{\rm LC}$ (G) & $1.0\times10^5$\\
    Energy flux\textsuperscript{a}, $F_{\gamma}$ (erg cm$^{-2}$ s$^{-1}$) & $8.363\times10^{-12}$\\
    Gamma-ray luminosity, $\mathcal{L}_{\gamma}$ (erg s$^{-1}$) & $4.0\times10^{33}$\\
    Gamma-ray efficiency, $\eta_{\gamma}$ & 14.3\%\\
    \hline
  \end{tabular}
  \vspace{2mm} 
    \parbox[l]{\linewidth}{\small
        \textsuperscript{a} Measured in the energy range between 100\,MeV and 100\,GeV, from 4FGL-DR4 \citep{4FGLDR4}.
    }
\end{table}

\begin{figure}
    \centering
	\includegraphics[width=\columnwidth]{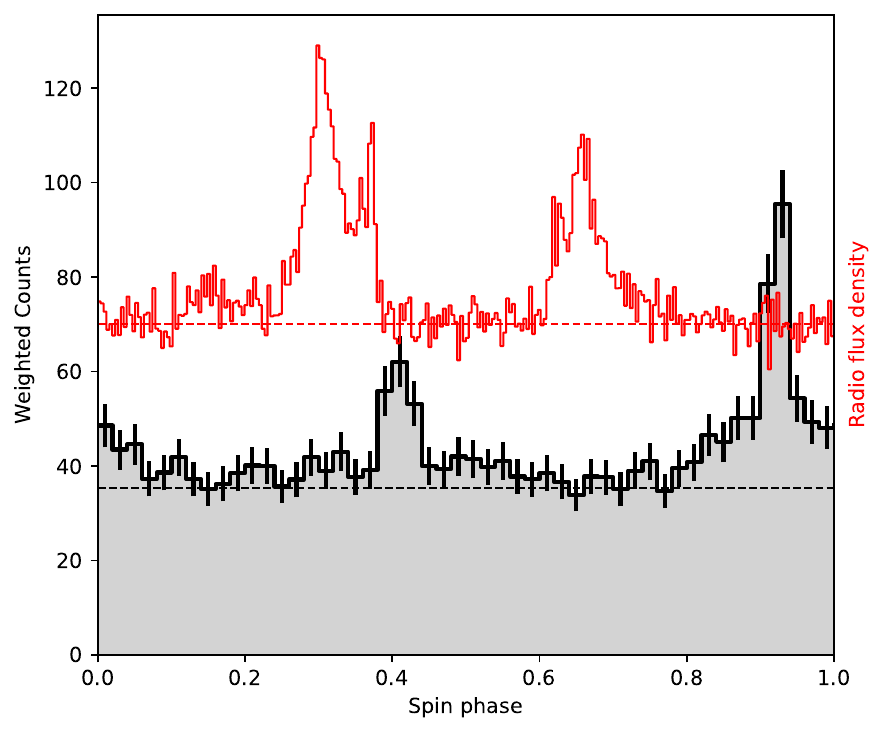}
    \caption{Phase-aligned radio (red) and gamma-ray (black) pulse profiles for PSR~J1544$-$2555. The radio profile based on MeerKAT UHF observations is shown in arbitrary flux density units from an arbitrary background level (dashed line).}
    \label{fig:align}
\end{figure}

In Figure~\ref{fig:align}, we show the phase-aligned radio and gamma-ray pulse profiles. The phase lag between radio and gamma-ray pulses is slightly ambiguous as the separation between the two gamma-ray peaks is roughly half a rotation apart, but looks comparable to the MSP population \citep{Smith2023,Burgay2024+timing}. A full alignment analysis could be used to estimate the binary inclination angle using emission models \citep[e.g., ][]{Corongiu2021+J2039}, but is outside the focus of this study.

\section{X-ray emission}
An X-ray source corresponding to PSR~J1544$-$2555 was first discovered by \citet{Mayer2024} and classified as a likely pulsar, in a systematic crossmatch between \textit{Fermi}-LAT and \textit{SRG}/eROSITA \citep{Predehl2021} sources. The detected X-ray emission is likely to be of nonthermal origin, either in the magnetosphere, or in an intrabinary shock of the system.  
While the X-ray source can be considered securely detected with a detection likelihood around 20, it has a relatively low count rate of $(3.3 \pm 0.9) \times 10^{-2}\, \rm ct/s$, in the main eROSITA energy band $0.2-2.3\,\rm keV$. This low count statistic prohibits a detection of X-ray pulsations in the eROSITA all-sky survey, as well as the execution of detailed spectral analysis.

We briefly estimate the intrinsic X-ray luminosity of the pulsar in the following way: the observed dispersion measure of $\mathrm{DM} = 25.817$ pc cm$^{-3}$ corresponds to an approximate X-ray absorption column density of $N_{\rm H} = 7.8\times 10^{20}\,\rm cm^{-2}$ \citep{He13}.
Assuming a power-law X-ray spectrum with a typical photon index of $\Gamma_{\rm X} = 2.0$, this corresponds to an unabsorbed source flux of $F = (8.0 \pm 2.2)\times 10^{-14}\,\rm erg\,s^{-1}\,cm^{-2}$ in the $0.5-8.0\,\rm keV$ band.
For the distance of $2.0\,\rm kpc$, obtained from the best-fit model of the optical light curve assuming a neutron star mass of 1.8 ${\rm M}_\odot$ (see section \ref{subsec:optical modelling}), this yields a pulsar luminosity of $L_{\rm X} \approx 4.0\times 10^{31} \,\rm erg\,s^{-1}$. This corresponds to an X-ray efficiency of $ L_{\rm X}/\dot{E} \approx 1.4\times 10^{-3}$, which is within the typical range for X-ray emitting pulsars \citep[e.g.][]{Becker1997}.
Similarly, the gamma-ray to X-ray luminosity ratio $L_{\gamma}/L_{\rm X} \approx 100$ is consistent with values commonly observed in MSPs \citep{Abdo2013}.

\section{Optical modelling}
\label{S:Optical_modelling}

\begin{figure*}
    \centering
    \includegraphics[width=\textwidth]{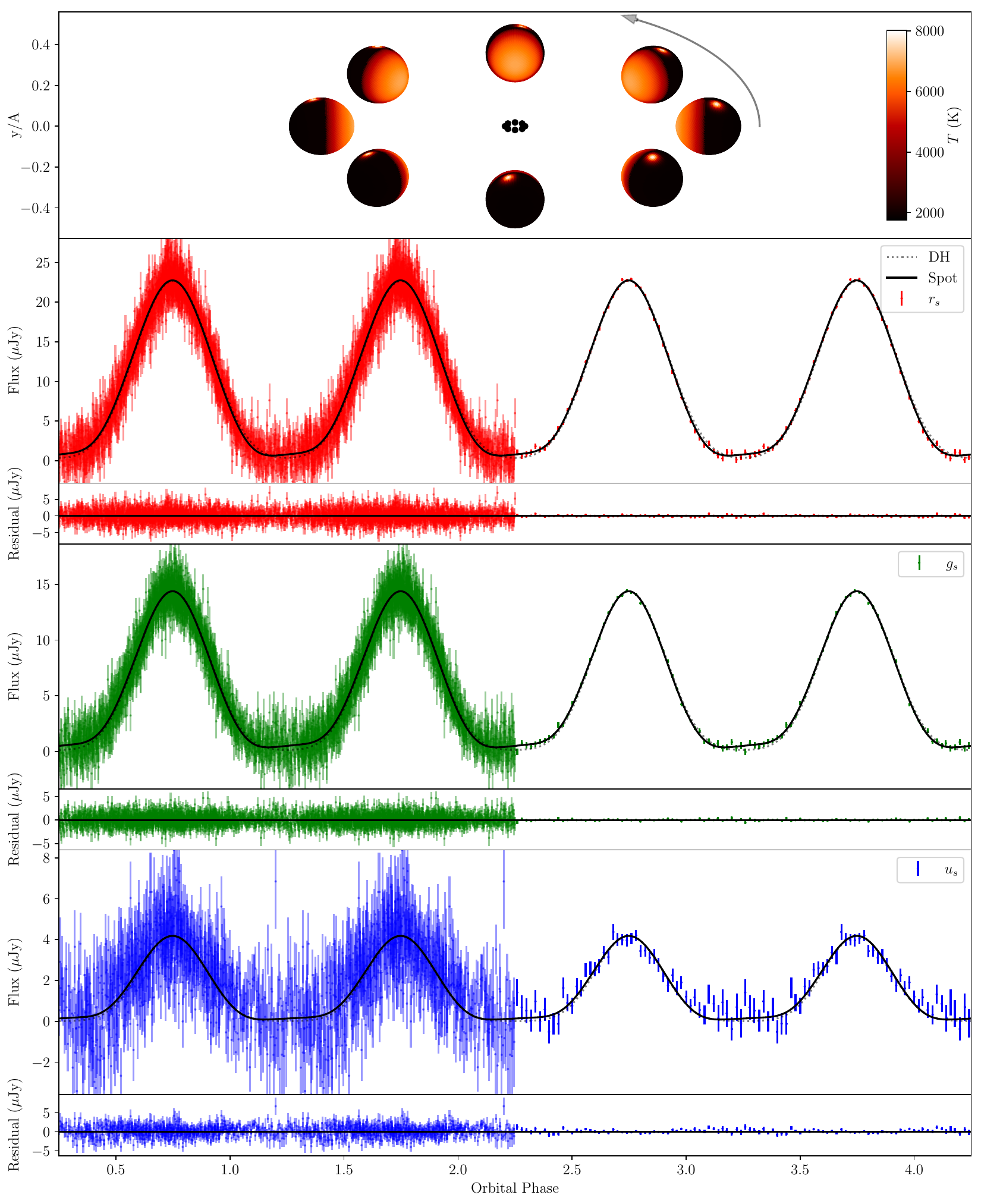}
    \caption{The schematic model of PSR~J1544$-$2555's companion with the spot model is shown in the uppermost panel. The subsequent panels display the light curves in $r_s$, $g_s$, and $u_s$, respectively. The observed data are presented for the first two cycles. These light curves are binned using an orbital phase interval of 0.02 and are shown in cycles 3 and 4.}
    \label{fig:j1544_lc_plot}
\end{figure*}

\subsection{Optical Modeling Configuration}
We used the stellar binary light curve synthesis code, \texttt{Icarus} \citep{Breton2011} to fit the light curves of PSR~J1544$-$2555. The code is parameterised by the mass ratio ($q$), orbital period ($P_{\rm orb}$), orbital inclination ($i$), rotational-to-orbital angular frequency ratio ($\omega$), projected radial velocity amplitude ($K_{\rm c}$), filling factor ($f_{\rm RL}$), gravity darkening exponent ($\beta$), base temperature ($T_{\rm base}$), irradiation temperature ($T_{\rm irr}$), V-band total extinction ($A_{\rm V}$) and distance modulus ($\mu$). $f_{\rm RL}$ is the ratio of the companion radius to the distance between the companion centre and the first Lagrangian point (L1). Modelled fluxes are evaluated from atmospheric grids which are generated from the atmospheric library, \texttt{ATLAS9} \citep{Castelli2004}. Full details about the procedure for creating atmospheric grids are described in \citet{Kennedy2022}. 

We assumed the companion to be tidally locked ($\omega$ = 1). The orbital period was obtained from the radio timing solution, and the projected semi-major axis ($x$) was used to derive other parameters in the binary mass function. Instead of fitting $A_{\rm V}$ and $\mu$, we fit the colour excess, $E(B-V)$, and distance ($d$). The $E(B-V)$ value is constrained using a Gaussian-tailed flat prior, with the upper tail peaking at $E(B-V)$ = 0.23 and a standard deviation of 0.02. These values are derived from the dust extinction map provided by \citet{Green2019}, assuming the binary lies at a distance between 0.5 and 6.5 kpc. The lower limit of the prior is set at $E(B-V)$ = 0, ensuring that the dust column density cannot drop below zero. This Gaussian-tailed flat prior approach accounts for the possibility that the observed dust column density might be lower than the upper limit due to the relatively low resolution of the dust map. Although the optical counterpart of PSR~J1544$-$2555 is associated with a star in the latest \textit{Gaia} DR3 release \citep{Gaia+DR3}, its parallax and distance have not yet been published. Therefore, we used the DM-derived distance of 1.02 kpc from the YMW16 model and the Galactic distribution of MSPs \citep{Levin2013} to form a prior distribution for the distance. Given that black widow companions are thought to be fully convective stars, we used a gravity darkening exponent of $\beta = 0.08$ \citep{Lucy1967}. Due to the large uncertainties in radial velocity measurements, we did not include them in the fitting along with the photometry. Instead, the observed spectra were compared with modeled spectra to verify consistency, particularly in the temperatures indicated by the prominent spectral lines. To improve parameter constraints, we fit the data with a fixed pulsar mass ranging from 0.6 to 3.6\,M$_{\odot}$, in increments of 0.2\,M${_\odot}$.

We initially fitted the observed light curves with the direct heating (DH) model where the irradiation incident on the companion is immediately reprocessed and re-emitted \citep{Breton2013}. However, these light curves exhibit slight asymmetries when compared to the modelled curves, as shown in Figure~\ref{fig:j1544_lc_plot}. To address these discrepancies, we also fitted the data using convection (C), diffusion-and-convection (D+C), and spot models. The C and D+C models are heat redistribution models that include both diffusion and convection terms, as described in \citet{Voisin2020}. The diffusion term is parameterised by the diffusion coefficient $\kappa$, while the convection term is parameterized by a coefficient $\nu_{\rm c}$ within a uniform angular velocity profile. The spot model is based on a model of high-energy particles directed toward the magnetic pole of its companion \citep{Sanchez2017}. Our spot model builds upon the direct heating model by introducing a Gaussian temperature profile centered at ($\phi_{\rm spot}$,$\theta_{\rm spot}$), with a radius $r_{\rm spot}$ and a temperature scaling factor $T_{\rm spot}$. Model optimisation was performed using \texttt{PyMultiNest} \citep{Buchner2014}, a Python wrapper for the \texttt{MultiNest} library \citep{Feroz2009}, which implements the nested sampling algorithm for Bayesian inference.

We calibrated the absolute flux of the observed light curves using $g$ and $r$ band data from the Pan-STARRS Data Release 2 \citep{PanSTARRS} and $u$ band data from the SkyMapper Southern Survey (SMSS) Data Release 4 \citep{SkyMapper}. Discrepancies in magnitude calibration between different instruments and filter systems could introduce small systematic deviations from the true flux of the star. To address this, we allowed the generated light curves to adjust within a Gaussian prior with a standard deviation of 0.05 magnitudes.

\begin{figure}
    \centering
    \includegraphics[width=\columnwidth]{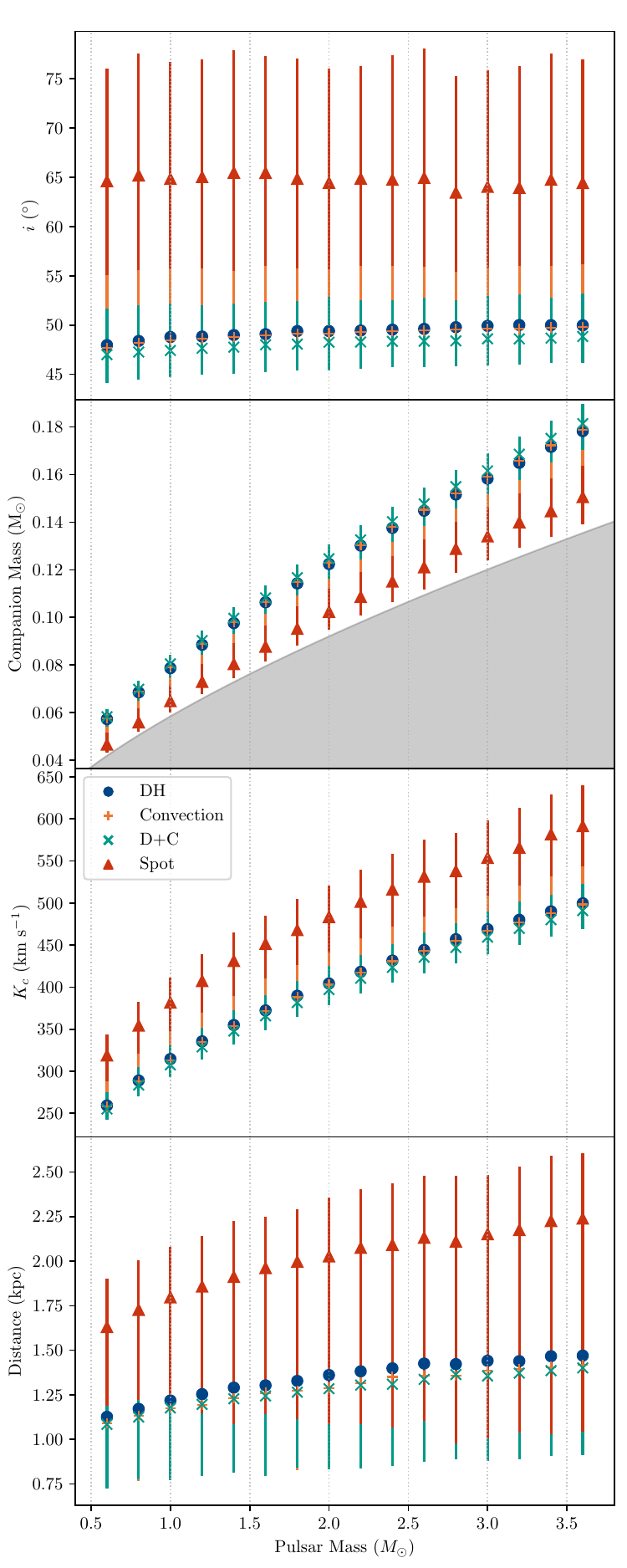}
    \caption{Parameter constraints of the four models against pulsar masses. The vertical bars represent the 95\% confidence intervals of posterior distributions. The shaded area denotes the forbidden parameter region ($i>90$\degr).}
    \label{fig:median_values_vs_mpsr}
\end{figure}

\subsection{Optical Modelling Results and Discussions}
\label{subsec:optical modelling}

\renewcommand{\arraystretch}{1.23}
\begin{table}
    \centering
    \caption{Fitted and derived parameters from the spot model after fixing $M_{\rm psr}$ to 1.8 $M_{\odot}$. $T_{\rm day}$ and $T_{\rm night}$ are the average temperatures of the companion's hemispheres facing toward and away from the pulsar, respectively. $T_{\rm dusk}$ and $T_{\rm dawn}$ are the average temperatures of the eastern and western hemispheres of the companion, respectively.}
    \label{tab:parameters}
    \begin{tabular}{lr}
    \hline
    Parameter & Value \\
    \hline
    \multicolumn{2}{c}{Fitted Parameters} \\
    \hline
    Colour Excess, E(B-V) & $0.21_{-0.08}^{+0.04}$ \\
    Companion radial velocity semi-amplitude, $K_{\rm c}$ (km s$^{-1}$) & $470_{-20}^{+20}$\\
    Distance, $d$ (kpc) & $2.0_{-0.3}^{+0.1}$ \\
    Orbital inclination, $i$ ($^\circ$) & $65_{-5}^{+6}$ \\
    Base temperature, $T_{\rm base}$ (K) & $1700_{-500}^{+600}$ \\
    Irradiation temperature, $T_{\rm irr}$ (K) & $6700_{-300}^{+300}$ \\
    Roche-lobe filling factor, $f_{\rm RL}$ & $0.65_{-0.1}^{+0.05}$ \\
    Spot longitude, $\phi_{\rm spot}$ ($^\circ$) & $-130_{-8}^{+10}$ \\
    Spot colatitude, $\theta_{\rm spot}$ ($^\circ$) & $30_{-10}^{+10}$ \\
    Angular radius of the spot, $R_{\rm spot}$ ($^\circ$) & $8_{-2}^{+3}$ \\
    Spot temperature, $T_{\rm spot}$ (K) & $6600_{-900}^{+1000}$ \\
    \hline
    \multicolumn{2}{c}{Derived Parameters} \\
    \hline
    Mass ratio, q & $18.9_{-0.9}^{+0.9}$ \\
    Companion mass, $M_{\rm c}$ (M$_\odot$) & $0.095_{-0.004}^{+0.005}$ \\
    Dusk-side temperature, $T_{\rm dusk}$ (K) & $4500_{-200}^{+200}$ \\
    Day-side temperature, $T_{\rm day}$ (K) & $6100_{-300}^{+200}$ \\
    Dawn-side temperature, $T_{\rm dawn}$ (K) & $4400_{-200}^{+200}$ \\
    Night-side temperature, $T_{\rm night}$ (K) & $2300_{-300}^{+300}$ \\
    Companion radius, $R_{\rm c}$ (R$_\odot$) & $0.171_{-0.02}^{+0.008}$ \\
    Volume-averaged filling factor, $f_{\rm VA}$ & $0.83_{-0.1}^{+0.05}$ \\
    \hline
    \end{tabular}
\end{table}

The parameter constraints in heat redistribution models are similar to those in the direct heating model, despite slightly better Bayesian evidence for the former. In contrast, the spot model exhibits a significantly improved goodness of fit and distinct parameter constraints compared to the other three models. Specifically, the spot model suggests an inclination angle of approximately 65\degr, consistent across different pulsar masses  (Figure~\ref{fig:median_values_vs_mpsr}), while the other models constrain the inclination to around 46–50 degrees. Due to the lack of precise radial velocity data, $K_{\rm c}$, the component masses could not be determined with high accuracy. Given that pulsar masses typically range between 1.0 and 2.1 ${\rm M}_\odot$, $K_{c}$ is likely between 400 and 500 km s$^{-1}$ for the spot model and 300–400 km s$^{-1}$ for the other models. The companion mass is estimated to be between 0.06 and 0.1 ${\rm M}_\odot$ for the spot model, and 0.08–0.12 ${\rm M}_\odot$ for the other models. Multi-band optical data, a large temperature contrast between the day and night sides of the companion, and precise radio ephemeris provide strong constraints on temperature parameters, resolving their degeneracy with distance and Roche-lobe filling. The fitted parameters for $T_{\rm base}$ and $T_{\rm irr}$ are approximately 1800 K and 6600–6700 K, respectively. Averaging the flux from the day and night hemispheres yields $T_{\rm day}\sim$ 6000–6200 K and $T_{\rm night}\sim$ 2300 K. The constrained distance range is 1.7–2.0 kpc, which exceeds the predicted DM distances from the YMW16 and NE2001 models. Table \ref{tab:parameters} reports the fitted and derived parameters from the spot model assuming a neutron star mass of 1.8 ${\rm M}_\odot$, corresponding to a value close to the median neutron star mass seen in redback systems \citep{Strader2019}.

\begin{figure}
    \centering
    \includegraphics[width=\columnwidth]{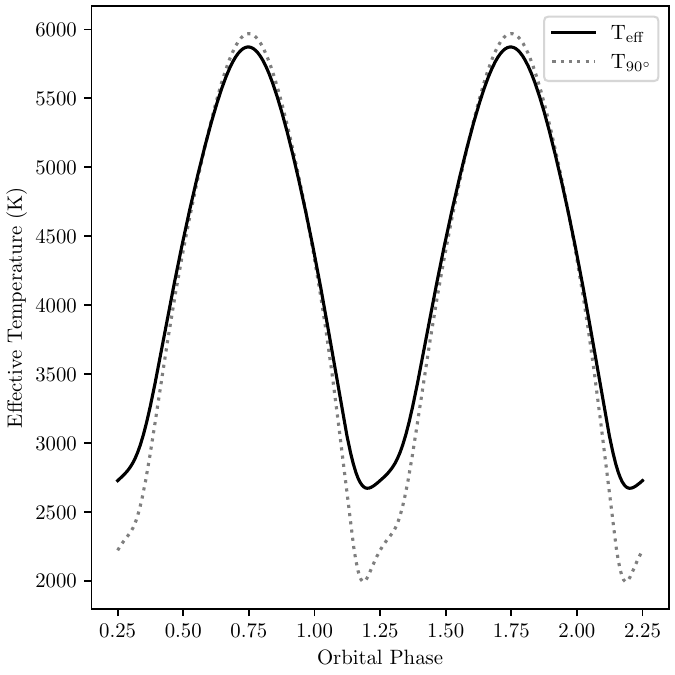}
    \caption{The flux-averaged temperature of the companion's hemisphere facing Earth is represented by the black solid line, indicating the effective temperature as inferred for the system's actual inclination. For comparison, the dotted gray line shows the flux-averaged temperature assuming the binary is viewed edge-on. This comparison highlights how the inclination angle affects the observed temperature distribution due to changes in the projected area and visibility of heated regions on the companion's surface.}
    \label{fig:companion_teff_plot}
\end{figure}

The spectra, taken between orbital phases 0.5 and 1.0 (near the light curve maximum), are dominated by noise. After smoothing them using a boxcar filter, we identified prominent lines such as Ca II K and H, CH, and weaker H Balmer lines, as shown in Figure~\ref{fig:j1544_spectra}. These lines are typical features of G-type stars whose effective temperatures are 5000-6000 K. The companion's effective average temperature from the spot model, as shown in Figure~\ref{fig:companion_teff_plot}, during the orbital phases between 0.5 and 1.0, also falls in the same temperature range.

The light curve near the maximum is less asymmetric compared to the orbital phases around the minimum, unlike other known black widows, where asymmetry is typically more pronounced near the peak. The light curve also shows a bluer colour near the minimum, as depicted in Figure~\ref{fig:j1544_lc_plot}. Of the four models fitted to the data, only the spot model accurately captures the observed light curves, particularly near the minimum. This model suggests a spot near the pole with a maximum temperature exceeding 8000 K and a width (corresponding to the standard deviation of the Gaussian spot profile) of about 8\degr.

A similar phenomenon, in which light curves appear bluer around the minima, was also observed in PSR~J1653$-$0158 \citep{Romani2014}. This effect is attributed to non-thermal emission, such as synchrotron radiation from intra-binary shocks. This emission makes the spectra appear bluer around the minima, when the companion's black-body radiation is at its faintest. The asymmetry in the light curve is likely due to the geometry of the shocks. This could also explain the similar effect observed in the companion of PSR~J1544$-$2555.

Although the spot model provides the best fit, no emission lines were detected that could originate from either a hot spot or an intra-binary shock. To distinguish between these two scenarios, one could fit a model combining direct heating with a non-thermal component, as implemented in \citet{Nieder2020}, or obtain X-ray spectra to check for synchrotron emission.

\subsection{Comparison with previous work}
When comparing our results with the modelling presented by \citet{Karpova2024+1544} (hereafter K24), several caveats must be considered. K24 adopted a direct heating model that does not account for gravity darkening. In addition, the quality and coverage of their light curve data differs from ours. Notably, their dataset lacks coverage around the orbital minimum, which significantly weakens constraints on $T_\text{base}$. The colour coverage is also non-uniform, with a paucity of data in the $V$ and $B$ bands. This limits the reconstruction of a consistent temperature profile across orbital phase and introduces degeneracies that propagate into key system parameters, such as the orbital inclination and Roche-lobe filling factor.

Despite these limitations, we perform a comparison between the K24 parameters and those derived from our best-fit model, transforming quantities as needed to match the \texttt{Icarus} framework. Following \citet{2019Zharikov}, we adopt the assumption $T_\text{night} = T_\text{base}$ and use the relation 

\begin{equation}
    T_\text{day} = T_\text{night} \times \left( 1 + \frac{F_\text{in}}{\Delta S \sigma T_\text{night}^{4}} \right)^{1/4} , \\
\end{equation}

where $\sigma$ denotes the Stefan-Boltzmann constant, and $F_\text{in}$ denotes the effective flux received from the pulsar by a surface element $\Delta S$ on the companion’s irradiated hemisphere. Using the maximum `day-side' temperature $T_\text{day,max}$ reported by K24, we compute the equivalent $T_\text{irr}$ in the \texttt{Icarus} context by subtracting the fourth power of $T_\text{night}$ from that of $T_\text{day,max}$. This value corresponds to $T_\text{irr}$ at the `nose' of the star, where the irradiation is maximal.

To compare with the parameter definition in \texttt{Icarus}, which evaluates $T_\text{irr}$ with respect to the companion’s centre of mass, we account for the geometry of the system, specifically the separation between the pulsar and companion, and the companion’s radius measured from its centre of mass toward the pulsar. Applying this correction yields an equivalent $T_\text{irr} \approx 6570$ K, consistent with our best-fit model. Other parameters such as the mass ratio, distance, and Roche lobe filling factor also show good agreement. 

However, the inclination and $T_\text{night}$ values reported by K24 are both higher than those found in our best-fit model. This is likely attributable to the absence of data near the orbital minimum in their light curve, which reduces sensitivity to the flux contribution from the night side. As a result, their modelling likely overestimates $T_\text{night}$, and to reproduce the observed modulation amplitude, a correspondingly higher inclination is required. This degeneracy between $T_\text{night}$ and inclination can be broken with multi-band observations such as those presented in the current work.

\section{Conclusions}

In this work, we have presented the discovery of a new black-widow millisecond pulsar, PSR~J1544-2555, associated with the \textit{Fermi}-LAT source 4FGL~J1544.2$-$2554. This discovery adds to the growing catalogue of known black widow pulsars and highlights the value of multiwavelength observations in the search for spider pulsar systems.

Optical observations conducted with ULTRACAM revealed a $\sim$2.7 hour orbital periodicity, consistent with black widow systems. This ephemeris enabled a targeted one-hour MeerKAT radio observation, optimised to avoid orbital phases prone to radio eclipses. We identified radio pulsations in an acceleration search on 15-minute segments, confirming the presence of a millisecond pulsar with a spin period of $\sim$2.4 ms. We derived a preliminary timing solution from a radio timing campaign with the Murriyang Parkes telescope, the Effelsberg 100-m radio telescope, and the Nançay Radio Telescope, which enabled us to detect gamma-ray pulsations in the \textit{Fermi}-LAT data. A full 16-year timing solution was obtained, allowing us to track orbital period variations. Furthermore, the associated X-ray source detected with eROSITA in the energy bands 0.2$-$2.3 keV and 2.3$-$5.0 keV, first discovered by \citet{Mayer2024}, though limited by the low count rates, provides additional confirmation of the source's non-thermal emission, characteristic of black widow systems.

Optical light curve modelling with \texttt{Icarus} favours a spot model, which effectively captured asymmetries around the minimum. This feature, along with the observed bluer colours at minimum, suggests potential non-thermal emission from intra-binary shocks. Further observations, including detailed X-ray spectroscopy, will be required to confirm this hypothesis. The quality of the spectra was insufficient for radial velocity measurements, preventing the determination of the neutron star’s mass. Future high-resolution spectroscopic studies could provide the radial velocity data needed to disentagle betweeen heating models, thus allowing to refine system parameters and ultimately measure the neutron star’s mass.

The discovery of PSR~J1544$-$2555 exemplifies the utility of the \textit{Fermi}-LAT catalogue in identifying MSP candidates and underscores the importance of optical surveys in detecting variable sources that can be effectively followed up through targeted radio observations.

\section*{Acknowledgements}

The authors would like to thank Soheb Mandhai and Pengyue Sun for their help with conducting optical observations.

The MeerKAT telescope is operated by the South African Radio Astronomy Observatory (SARAO), which is a facility of the National Research Foundation, an agency of the Department of Science and Innovation. We thank staff at SARAO for their help with observations and commissioning. TRAPUM observations used the FBFUSE and APSUSE computing clusters for data acquisition, storage and analysis. These clusters were funded and installed by the Max-Planck-Institut f\"{u}r Radioastronomie (MPIfR) and the Max-Planck-Gesellschaft. The National Radio Astronomy Observatory is a facility of the National Science Foundation operated under cooperative agreement by Associated Universities, Inc. The Parkes radio telescope is part of the Australia Telescope National Facility (https://ror.org/05qajvd42) which is funded by the Australian Government for operation as a National Facility managed by CSIRO. We acknowledge the Wiradjuri people as the traditional owners of the Observatory site. The Nan\c cay Radio Observatory is operated by the Paris Observatory, associated with the French Centre National de la Recherche Scientifique (CNRS) and Universit\'{e} d'Orl\'{e}ans. It is partially supported by the Region Centre Val de Loire in France.

The \textit{Fermi} LAT Collaboration acknowledges generous ongoing support from a number of agencies and institutes that have supported both the development and the operation of the  LAT  as  well  as  scientific  data  analysis.  These  include  the National  Aeronautics  and Space   Administration   and   the   Department   of   Energy   in the   United   States,   the Commissariat \`{a} l'Energie Atomique and the Centre National de la Recherche Scientifique /  Institut  National  de  Physique  Nucl\'{e}aire et  de  Physique  des  Particules  in  France,  the Agenzia  Spaziale  Italiana and the Istituto  Nazionale  di  Fisica  Nucleare  in  Italy,  the Ministry  of  Education,  Culture, Sports,  Science  and  Technology  (MEXT),  High  Energy Accelerator  Research Organization  (KEK)  and  Japan  Aerospace  Exploration  Agency (JAXA)  in  Japan, and  the  K.  A.  Wallenberg  Foundation,  the  Swedish  Research  Council and the Swedish National Space Board in Sweden. 

This work has made use of data from the European Space Agency (ESA) mission {\it Gaia} (\url{https://www.cosmos.esa.int/gaia}), processed by the {\it Gaia} Data Processing and Analysis Consortium (DPAC, \url{https://www.cosmos.esa.int/web/gaia/dpac/consortium}). Funding for the DPAC has been provided by national institutions, in particular the institutions participating in the {\it Gaia} Multilateral Agreement.

T.T. is grateful to the National Astronomical Research Institute of Thailand (NARIT) for awarding a student scholarship. R.P.B. acknowledges support from the European Research Council (ERC) under the European Union's Horizon 2020 research and innovation program (grant agreement No. 715051; Spiders). SBD acknowledges the support of a Science and Technology Facilities Council (STFC) stipend (grant number: ST/X001229/1) to permit work as a postgraduate researcher.
E.C.F. is supported by NASA under award number 80GSFC24M0006. ULTRACAM and VSD are supported by STFC. Lastly, we would like to thank Lucas Guillemot for reviewing the manuscript on behalf of the \textit{Fermi}-LAT Collaboration.

\section*{Data Availability}

TRAPUM and ULTRACAM observations are available upon reasonable request to the contact author. Reduced and calibrated optical light curves are available in accompanying online material. The \textit{Fermi}-LAT data are available from the \textit{Fermi} Science Support Center (FSSC, \url{http://fermi.gsfc.nasa.gov/ssc}). Gamma-ray timing solutions are also available from the FSSC (\url{https://fermi.gsfc.nasa.gov/ssc/data/access/lat/ephems/}).



\bibliographystyle{mnras}




\appendix

\section{The Modelled Spectra and the Posterior Distributions of the Optical Modelling}

\begin{figure*}
    \centering
    \includegraphics[width=\textwidth]{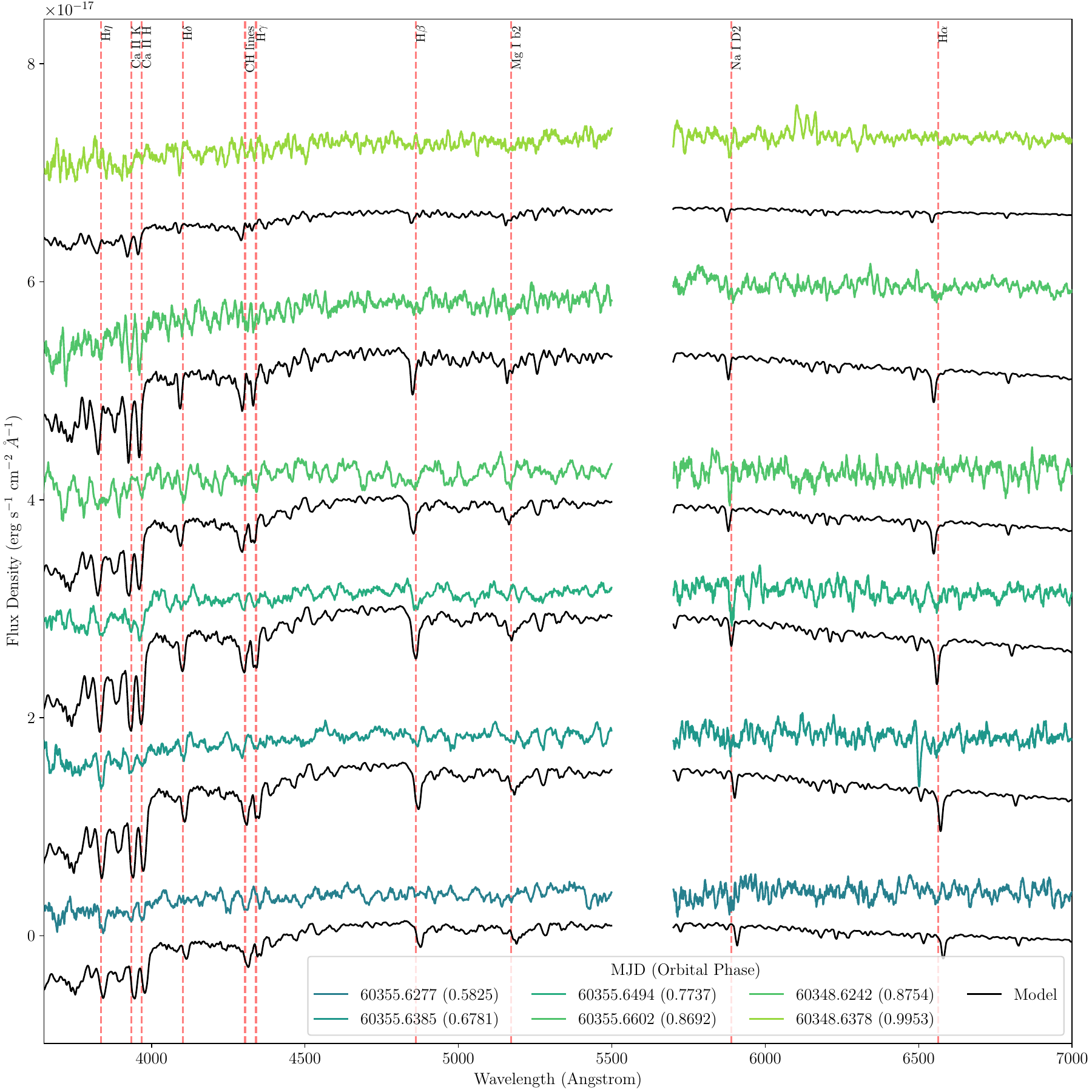}
    \caption{Observed spectra are plotted with an offset between them. The simulated spectra from the spot model using \texttt{Icarus} are shown below its counterpart observed spectra. Both types of spectra are convolved with a boxcar filter (width = 9) to smooth out the noise and see lines better. Notable absorption lines are labelled at the top of the plot.}
    \label{fig:j1544_spectra}
\end{figure*}

\begin{figure*}
    \centering
    \includegraphics[width=\textwidth]{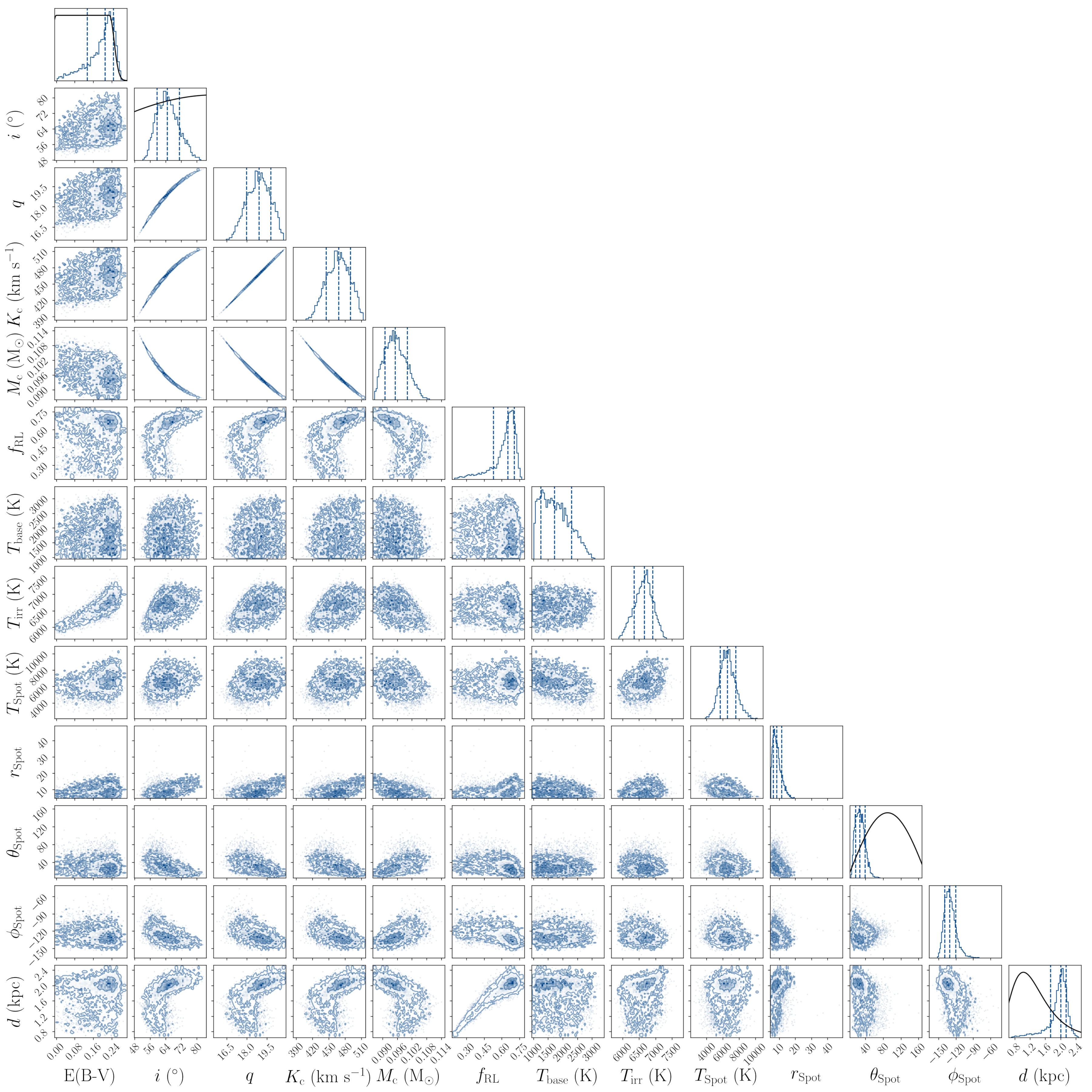}
    \caption{Corner plots displaying the various posterior distributions of the spot model where pulsar mass is fixed to 1.8 M$_{\odot}$. The black curves in the 1D distributions show the priors distributions that were adopted.}
    \label{fig:j1544_optical_corner_plot}
\end{figure*}


\bsp	
\label{lastpage}
\end{document}